\documentstyle[aps,pre]{revtex}
\input{psfig.tex}

\begin{document}

\draft

\title{Anomalous scaling behavior in Takens-Bogdanov bifurcations}

\author{E.~R.~Tracy\footnote[1]{Email:tracy@rayleigh.physics.wm.edu}}
\address{Physics Department, College of Wm. \& Mary, Williamsburg, VA
23187-8795}

\author{X.~Z.~Tang\footnote[2]{Email:tang@chaos.ap.columbia.edu}}
\address{Department of Applied Physics, Columbia University,
New York, NY 10027}

\date{October 22, 1997}

\maketitle

\begin{abstract}

A general algorithm is presented for estimating the nonlinear instability
threshold, $\sigma_c$, for subcritical transitions in systems where the
linearized dynamics is significantly non-normal (i.e. subcritical
bifurcations of {\em Takens-Bogdanov} type).  The $N$-dimensional 
degenerate node is presented as an example.
The predictions are then compared to numerical studies with 
excellent agreement.

\end{abstract}

\pacs{PACS numbers:  47.20.Ky, 47.20.Ft, 47.27.Cn}
\vspace{0.5cm}


Consider a {\em nonlinear} dynamical system whose dynamics in the vicinity of 
a stable equilibrium is non-normal
at {\em linear} order.  (A matrix or linear operator, ${\cal L}$, is 
non-normal if ${\cal L}^{\dag}{\cal L}\neq 
{\cal L}{\cal L}^{\dag}$ with ${\cal L}^{\dag}$ the 
adjoint.)  In the subcritical case,
nonlinearity makes the equilibrium unstable to finite-amplitude
perturbations.  The goal is to estimate the size of the smallest impulse 
needed to drive the system unstable, denoted $\sigma _c$.  Non-normality
implies that the eigenvectors of ${\cal L}$ are not orthogonal.  In
the extreme case of degeneracy two, or more, of the eigenvectors can
become parallel leading to non-diagonalizability of the operator (see,
for example,\cite{golub} for a discussion of the finite dimensional case).
A prototypical example in two dimensions is that studied by 
Takens~\cite{takens}
and Bogdanov~\cite{bogdanov}(here the ${\dot x}_1=dx_1/dt$ etc.):\\
\begin{equation}
\left(
\begin{array}{c}
{\dot x}_1\\
{\dot x}_2
\end{array}
\right)=
\left(
\begin{array}{cc}
0 & 1\\
0 & 0
\end{array}
\right)
\left(
\begin{array}{c}
x_1\\
x_2
\end{array}
\right)+
\left(
\begin{array}{c}
f_1(x_1,x_2)\\
f_2(x_1,x_2)
\end{array}
\right)
\end{equation}
\vspace*{.1in}\\
with $f_1$ and $f_2$ nonlinear functions of their arguments.
In a slight abuse of terminology, in the following discussion we will use the terms 
`Takens-Bogdanov bifurcations' and `non-normal transitions' interchangeably.
Hence, a `Takens-Bogdanov bifurcation' or a `non-normal transition' here shall
mean any system where the linearized dynamics is non-normal and the
non-normality produces important effects.

The notion that non-normality might be physically important was first
suggested by Orr~\cite{orr} in an attempt to explain the failure of standard linear
stability analysis to predict the observed critical Reynold's 
number for the laminar/turbulent transition in some shear flows.
This conjecture has been revived recently
(see~\cite{trefethen,baggett1,boberg,farrell1,grossman} for 
pro and \cite{waleffe1} for con).
Such degenerate transitions occur in other physical models as well. For
example they arise in aeroelastic models~\cite{holmes}, and stall 
models for turbines~\cite{mccaughan}.  For a discussion of other physical
applications the interested reader is referred to Chapter 7 of the most recent 
edition of~\cite{guckenheimer}.  The related degenerate Hopf is discussed 
in~\cite{crawford}.

Much of the work cited above deals with bifurcations of low-dimensional models 
derived from more primitive equations (for example Navier-Stokes) by Galerkin 
projection.  When such projections exhibit degeneracies one must be careful
about the physical interpretation.  Unless one can demonstrate that the
behavior of interest is robust under perturbation then it is most likely that the
degeneracy is a mathematical pathology with little physical importance.

Such perturbations come in many varieties, with two of most important being:  
a) perturbations of the equations themselves (meaning, for example, the degeneracy 
is not exact), and b) noise driving, both additive and multiplicative. Such noise might 
represent, for example, coupling to the environment or to degrees of freedom which have 
been projected out by the Galerkin procedure.
In this letter we consider perturbations of type a).  The additive noise response
of Takens-Bogdanov systems will be discussed elsewhere~\cite{noisy}.

A key question from a physical point of view is:  {\em What robust observable 
characteristics, if any, distinguish `non-normal' transitions from `normal' ones?} 

Baggett and Trefethen~\cite{baggett1} have shown numerically that a range of 
low-dimensional non-normal systems exhibit 
anomalous scaling behavior at subcritical transitions:  the rate at which 
$\sigma _c \downarrow 0$ as the threshold for linear instability is 
approached differs markedly from that of normal systems.  This is 
summarized by the scaling law $\sigma _c\sim \epsilon ^{\gamma}$ 
with $\epsilon $ the linear stability parameter.  (Here $\epsilon$ 
is used instead of the inverse Reynolds' number, $1/R$.  The 
threshold is $\epsilon =0$ with $\epsilon >0$ stable.)
For normal transitions $\gamma $ is generically unity, while for non-normal 
transitions $\gamma $ can be greater than unity. 
Baggett has also shown that it is possible to derive the anomalous
scaling exponent in a simple $2$-dimensional case~\cite{bagthesis}.

The purpose of this letter is to provide a {\em general algorithm} for computing 
the scaling exponents and to illustrate the geometric origin of the anomalous
scaling behavior.  The basic logic 
parallels that of threshold estimation for normal subcritical systems:  

First,  at the linear instability threshold ($\epsilon =0$), a normal form 
analysis~\cite{guckenheimer} identifies
nonlinear terms which cannot be removed -- {\em i.e. } transformed to higher 
nonlinear order -- by coordinate transformations.  These are called 
{\em resonant} nonlinear terms.  Here, the linear term is assumed to be 
{\em exactly} degenerate and non-diagonalizable.

Second, backing away from the linear threshold ($\epsilon >0$), an
{\em asymptotic} analysis is performed on the resonant nonlinear terms  
to determine which of them dominate.  Here, the degeneracy could be weakly
broken as well.  The physical intuition is that the most important nonlinear 
terms are both {\em resonant} (in the normal form) and 
{\em dominant} (in the asymptotic limit of $\epsilon \downarrow 0$).
As will be shown, for the $2$-dimensional degenerate node this combined
normal form/asymptotic balance method identifies the same dominant nonlinear
term as reported by Takens~\cite{takens}.  Takens, however, used the technique
of {\em topological blowup} to identify the dominant nonlinear term,
requiring {\em three} blowups in sequence before the dominant term was 
revealed~\cite{guckenheimer}.

While the method reported here is quite general, it is illustrated on an 
$N$-dimensional degenerate node because this is relatively simple.
The extension to general $N$ is non-trivial and is necessary for physical 
applications where more than $2$ degrees of freedom may be near threshold.
To our knowledge, the topological blowup analysis of Takens has not been
extended to higher dimensions.  

Before starting the analysis, we state our main conclusion: the anomalous
scaling behavior of non-normal transitions is determined 
purely by an appropriate balance between the dominant linear and nonlinear effects,
as identified by the normal form/asymptotic balance analysis. 
The new element in the non-normal balance is a geometric relation, summarized
by the {\em triangle relation}, Figure~(1).  We turn now to the presentation of
the algorithm.  We consider the discrete-time
map as well as the continuous-time flow to illustrate the simplicity of the result.
The flow is gotten by an appropriate limit of the map.

\begin{figure}
\centerline{\psfig{figure=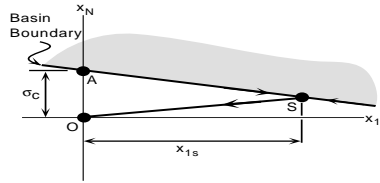,height=2.5in,width=2.5in}}
\caption{
The asymptotic scaling relation between $x_{1s}$ and $\sigma _c$ is 
summarized by the {\em triangle relation} of this figure.  The triangle
OSA is formed by the node (O), the saddle (S) and the point of closest
approach of the basin boundary to the node (which lies in the immediate
neighborhood of (A)).  The shape of this triangle
is determined by the {\em linear} amplification factor 
({\protect{\ref{scaling2}}}):
$x_{1s}/\sigma _c:|x_1|_{\max}/\sigma _0$.}
\label{figure:triangle}
\end{figure}

Consider the discrete-time dynamical system:
\begin{equation}
x_j(m+1)=F_j(x(m);p)\qquad m=0,1,2\ldots
\end{equation}
where $x, F\in \Re ^N$, $F$ is a smooth nonlinear function of $x$ and $p$ is a 
control parameter. Suppose $F$ has a fixed point $x_*(p)$ which we take to be the
origin.  Expanding to first order:
\begin{equation}
x_j(m+1)=\sum _{k=1}^{N} A_{jk}x_k(m);\qquad j=1,2\ldots N,
\label{map}
\end{equation}
with $A\equiv \nabla F|_{x=0}$.  We take $A$ to be of the form:
\begin{equation}
A\equiv 
\left(
\begin{array}{cccc}
\lambda    & \alpha   & \ldots &   0        \\
  0        & \lambda  & \alpha & \vdots     \\
\vdots     &          & \ddots & \alpha     \\
  0        & \ldots   &        & \lambda
\end{array}
\right).
\label{linear_matrix}
\end{equation}
All off-diagonal terms (except $\alpha$) are zero. The eigenvalue
$\lambda $ is real and $0<\lambda \leq 1$.  The coupling constant, $\alpha$, 
is assumed to be real and positive.  More general $A$ will be treated in a 
longer paper.

The map~(\ref{map}) can be considered an Euler integrator for the flow
${\dot x}=(A-I)x$ if we take $\lambda -1\sim \delta t$ and 
$\alpha \sim \delta t$, with $\delta t$ the step-size.   Hence, 
the discrete-time node will have the same scalings as the related
flow.  In fact, although the scaling exponent (\ref{scaling3}) is developed 
using maps, the numerical tests (Figure (2)) were all performed using flows.  
The results are in complete agreement.

\begin{figure}
\centerline{\psfig{figure=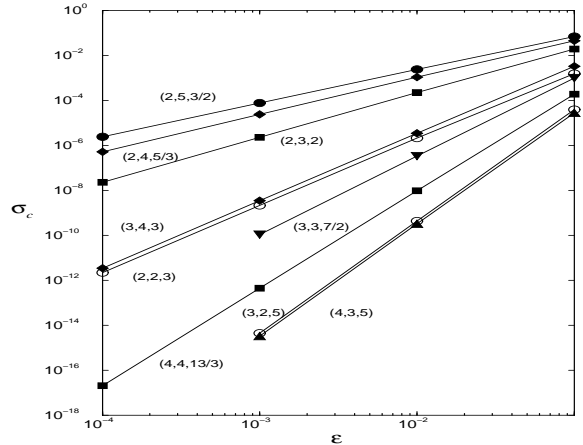,height=2.5in,width=2.5in}}
\caption{The scaling exponents $\gamma(N,n)$ are tested for various 
combinations of $N$ (the dimension of the dynamics) and $n$ (the leading 
order nonlinearity).  The symbols are the numerical results which were
generated using a Bulirsch-Stoer method {\protect\cite{numrec}} and the
lines are the predicted scalings.  The results are labeled as $(N,n,\gamma).$}
\label{figure:scaling}
\end{figure}

Consider the {\em linear} impulse response.  The analysis for the flow is
straightforward as one is dealing with an $N$-dimensional system of equations
with constant coefficients.  Hence $x(t)=exp(At)x_0$.  The degeneracy of $A$
implies that the components of $x(t)$ will not be simple exponents in $t$, but
of the form $t^ne^{-\gamma t}$ for some $\gamma$.  This result can be found in
any undergraduate text on differential equations.  
The calculation for the discrete-time map is more challenging:

The system is given a
random initial kick at $m=0$ with correlation matrix:
$<x_0x^T_0>=\sigma ^2_{0} I$
and $I$ the $N\times N$ identity matrix.
At the $m$th time step $x(m)=A^mx_0$.  If $A$ is banded, then so is
$A^m$, hence one need only calculate the last column, $[A^m]_{jN}$:
\begin{equation}
[A^m]_{jN}= \left(
\begin{array}{c}
m \\
N-j
\end{array}
 \right)\lambda ^m
\left({\alpha \over \lambda }
\right)^{N-j},
\end{equation}
with $
\left(
\begin{array}{c}
m \\
N-j\end{array}
 \right)$
the binary coefficients.  The index $j$ ranges from $N-m$ to $N$ for
$m<N$ (with all entries above the $j=N-m$ entry still zero), and from $1$ to 
$N$ when $m\geq N$.

The norm 
$||x(m)||^2=x(m)\cdot x(m)=x_0\cdot [A^m]^TA^mx_0$.
Taking the ensemble average over the initial conditions gives:
$<||x(m)||^2>=\sigma ^2_{0}Tr\left([A^T]^mA^m\right)$.
For $m>>N-j$ the trace of $[A^T]^mA^m$ is dominated by the 
contribution from $[A^m]_{1N}$, hence we take that as our estimate:
\begin{equation}
<||x(m)||^2>^{1/2}\sim |x_1(m)|\sim \sigma _{0}\lambda ^m
\left(
\begin{array}{c}
m \\
N-1
\end{array}
\right)
\left(
{\alpha \over \lambda}
\right)^{N-1}
\label{norm}
\end{equation}

The above result holds for arbitrary $\lambda $ and $\alpha$ as long as
$\alpha > {\cal O}(\epsilon)$.  Now assume $\lambda =1-\epsilon$ and consider
$\epsilon \downarrow 0$.  Taking $m\sim t/\epsilon $ with 
$t\sim {\cal O}(1)$ and using Stirling's formula 
$m!\sim m^me^{-m}(2\pi m)^{1/2},$ one has
\begin{equation}
|x_1(t)|\sim \sigma _{0}{1\over {(N-1)!}}
\left({{\alpha} \over \epsilon }
\right)^{(N-1)}
t^{N-1}e^{-t}
\end{equation}
As a function of $t$, this reaches a maximum when $t=N-1$, therefore:
\begin{equation}
{{|x_1|_{max}}\over {\sigma _0}}\sim \left({{\alpha} \over \epsilon }
\right)^{(N-1)}
\label{scaling2}
\end{equation}
where we have suppressed the $N$-dependent prefactor.  This result is the 
amplification factor of the linear transients (note that it is trivially 
valid if $N=1$).

Now consider the effects of nonlinearity, and treat first the $N=2$ case
for simplicity.  We assume only that the nonlinear terms are smooth.
To study generic behavior one casts the given problem into its simplest form
by performing smooth and invertible coordinate transformations to eliminate
as many nonlinear terms as possible.  The normal form analysis identifies
those terms which {\em cannot} be eliminated.  This
analysis is done for $\epsilon =0$, then $\epsilon > 0$ reintroduced for the
asymptotic balance estimates to follow.  For a detailed discussion of
the Takens-Bogdanov normal form analysis, the interested reader is refered
to Chapter 7 of~\cite{guckenheimer}.

To quadratic order, the normal form for the $N=2$ degenerate node is
\begin{equation}
\begin{array}{l}
x_1'=(1-\epsilon)x_1+\alpha x_2 \\
x_2'=(1-\epsilon)x_2+a_2x_1x_2+b_2x_1^2
\end{array}
\label{quad}
\end{equation}
with $a_2$ and $b_2$ arbitrary coefficients.  If these quadratic terms
do not appear, then one must go to higher order (all other quadratic terms
can be pushed to higher order by changing coordinates).  At $n^{th}$ order one 
finds
\begin{equation}
\begin{array}{l}
x_1'=(1-\epsilon)x_1+\alpha x_2 \\
x_2'=(1-\epsilon)x_2+a_nx_1^{n-1}x_2+b_nx_1^n.
\end{array}
\label{cubic}
\end{equation}
For concreteness assume $a_n$ and $b_n$ are positive (or 0).  This insures
the system is subcritical.  Consider $n=2$: 
solving~(\ref{quad}) for the position of the saddle (find the second root of 
$x'=x$):
$\epsilon x_{1s}=\alpha x_{2s};\quad
\epsilon x_{2s}=a_2x_{1s}x_{2s}+b_2x_{1s}^2$.
There are two simple cases:\\

\noindent{(I):  $a_2= 0, b_2\sim {\cal O}(1)$;}
$\;\; x_{1s}= \epsilon ^2/\alpha b_2,\;\; x_{2s}=(\epsilon /\alpha)x_{1s}$.\\

\noindent{(II): $a_2\sim {\cal O}(1), b_2= 0$;}
$\quad x_{1s}=\epsilon /2a_2,\;\; x_{2s}=0.$\\

Rescaling via $x_{1s}\equiv \epsilon ^2{\tilde x}_{1s}$ and
$x_{2s}\equiv \epsilon ^3{\tilde x}_{2s}$ reveals that the
$b_2$ terms dominates.  Hence, the scaling for Case I should
be seen most often for quadratically nonlinear systems, while that of Case 
II requires special conditions (the smallness of $b_2$ to at least 
${\cal O}(\epsilon)$).  ({\em N.B.} As mentioned earlier, this identification of 
$b_2$ as the dominant term agrees with the topological blowup analysis of 
Takens~\cite{takens,guckenheimer}.  The approach we describe in this
letter differs from a blowup analysis in that we are considering asymptotic
balances in the neighborhood of the bifurcation ($\epsilon >0$) while the
blowup analysis is done at the bifurcation ($\epsilon =0$).)

At cubic order, it is the $b_3x_1^3$ term driving $x_2$ which
dominates.  This behavior holds for general $nth$-order nonlinearities,
which allows us to state:  if the first nonlinearities appear at order
$n$, and $b_n>0$ and $\sim{\cal O}(1)$, then the position of the saddle
is given by 
$\epsilon x_{1s}\sim\alpha x_{2s};\quad \epsilon x_{2s}\sim b_nx_{1s}^n.$
This implies $x_{1s}\sim 
\left(
{\epsilon \over {\alpha b_n}}
\right)^{1/(n-1)};
\quad
x_{2s}\sim \left(b_n\over \epsilon \right)x_{1s}^n.$

In higher dimensions ($N>2$), the normal form analysis reveals that new 
resonances become possible (with one new resonance appearing for each
increment $N\rightarrow N+1$).  Most importantly, the
$b_nx_1^n$ term will {\em always} resonantly drive $x_N$.  Asymptotic 
estimates show that if $b_n>0$ and  $\sim {\cal O}(1)$ this term 
will be dominant.  The position of the saddle is:
\begin{equation}
\epsilon x_{1s}\sim \alpha x_{2s};\;\ldots 
\epsilon x_{kx}\sim \alpha x_{k+1};\;\ldots
\epsilon x_{Ns}\sim b_nx_{1s}^n,
\end{equation} 
which gives
\begin{equation}
x_{1s}\sim 
\left[
{\epsilon \over b_n}
\left(
{\epsilon \over \alpha }
\right)^{N-1}
\right]^{1/(n-1)}
\label{saddle}
\end{equation}
Note that this estimate is also valid in the normal case where, because
the linear term is diagonalizable, $N$ is effectively 1.

We now turn to the estimate of the subcritical threshold:  as $\epsilon \downarrow 0$,
how far is the basin boundary from the degenerate node?
Eq.(\ref{saddle}) gives the distance to the basin boundary {\em along the 
stable manifold of the node}.  If this were
a {\em normal} saddle-node bifurcation, $x_{1s}$ 
would typically give a good estimate of the distance of closest approach of 
the basin boundary.  However, the {\em non-normal} linear behavior
forces the basin boundary to form an acute angle with the stable manifold
of the node, hence it will lie very close to the node in 
directions transverse to the stable manifold.  This is summarized by the
triangle relation of Figure~(\ref{figure:triangle}).  The linear response 
determines the shape of the triangle and relates $|x_1|_{max}$ to an initial
impulse $x_{N0}\sim \sigma _0$ via Eq. (\ref{scaling2}), i.e.
$|x_1|_{max}\sim (\alpha /\epsilon)^{N-1}\sigma _0$.  If 
$|x_1|_{max}\sim |x|_{1s}$, then this initial perturbation will have crossed
the basin boundary.  Therefore, this balance determines the threshold for 
instabilities due to finite perturbations and gives as a threshold estimate:
\begin{equation}
\sigma _c\sim (\epsilon /\alpha)^{N-1}x_{1s}
=
{\epsilon ^{\gamma }\over {b_n\alpha ^{\gamma -1}}}
\end{equation}
with 
\begin{equation}
\gamma (N,n)\equiv 
{{n(N-1)+1}\over {n-1}}
\label{scaling3}
\end{equation}
Eq.(\ref{scaling3}) is our primary result.  We note that this is also valid
for normal ($N=1$) case.

Table~\ref{table:exponent} summarizes the scaling exponents, 
$\gamma $, for several $N$ and $n$.  
These were tested numerically with {\em flows}. The flows have 
linear dynamics $\dot{x}=A x$ with $A$ of the form~(\ref{linear_matrix}) 
and $\lambda$ replaced by $-\epsilon.$
The models all have the dominant nonlinear term $x_1^n$ driving 
the $x_N$ component.  All nonlinear coefficients were set to unity. 
The initial conditions were set to be $(0,\cdots,x_N^0).$ 
The scaling exponents were computed by plotting
the critical $x_N^0$ beyond which trajectories escape to $||x||>>|x_{1s}|$.
The results are summarized in
Figure (2).  As can be seen, the observed scalings agree completely with
the prediction~(\ref{scaling3}).\\

In summary, we have shown that it is possible to systematically evaluate the 
importance of various nonlinear effects on non-normal transitional behavior
by an extension  of the techniques used for normal systems.  This leads
to an algorithm capable of predicting the nonlinear threshold for subcritical 
transitions.  The algorithm was illustrated by application to an
$N$-dimensional degenerate node, where the anomalous scaling behavior was
shown to be due to the fact that the non-normality of the linear term 
forces a geometrical relationship between the length scales along and across 
the stable manifold of the node, an effect which is absent in normal systems. 
The resulting scaling exponent~(\ref{scaling3}) shows 
that non-normality and nonlinearity act together to increase the sensitivity 
to subcritical transitions, and the threshold depends exponentially
on the number of degrees of freedom taking part, $N$.  Figure
(\ref{figure:scaling}) suggests that by measuring such scaling behavior near
threshold it might be possible to choose between various models (or at least
eliminate a large class of them), though if $N$ is large such scaling
regimes will be extremely narrow in $\epsilon$.\\

\noindent{We thank J. S. Baggett for useful comments.
This work was supported by the AFOSR and the DOE.}

\begin{table}
\begin{tabular}{ccccc}
{\protect{$\gamma (N,n)$}}    & n=2  & n=3   & n=4  & n=5 \\ 
\hline
N=2             & 3*   & 2**   & 5/3  & 3/2 \\
N=3             & 5    & 7/2   &  3   & 11/4  \\
N=4		& 7    & 5     &  13/3 & 4  
\end{tabular}
\caption{Tabular summary of {\protect{$\gamma $}} 
computed from ({\protect\ref{scaling3}}). 
The highlighted entries correspond to models in {\protect\cite{baggett1}}, 
with (their notation) {\protect{*=TTRD', **=TTRD''.}}  As mentioned in
the text, although both of their models are nominally quadratic, the
normal forms are quite different and show that {\protect{TTRD''}} can in fact 
be transformed to be cubically nonlinear.  Their reported threshold scalings 
for {\protect{TTRD'}} and {\protect{TTRD''}} are $3$ and $2$, respectively.  
These are the only two models
we compare with {\protect\cite{baggett1}} because the rest of their models
either have a square root singularity at the origin -- hence the normal form
analysis does not apply -- or the 
models are not uniform in their coupling, implying the normal 
form used here would not be the correct one.}
\label{table:exponent}
\end{table}


\begin{thebibliography}{99}

\bibitem{golub} G. H. Golub \& C. F. Van Loan, {\em Matrix computations, $2^{nd}$
edition}, (Johns Hopkins, Baltimore, 1991).

\bibitem{takens} F. Takens, Publ. Math. I. H. E. S. {\bf 43}, 47 (1974).

\bibitem{bogdanov} R. I. Bogdanov, Func. Anal. Appl. {\bf 9}, 144 (1975).

\bibitem{orr} W. M'F. Orr, Proc. R. Irish Acad. Ser. A {\bf 27}, 9 (1907).

\bibitem{trefethen} L. N. Trefethen, A. E. Trefethen, S. C. Reddy, and
T. A. Driscoll, Science {\bf 261}, 578 (1993).

\bibitem{baggett1}  J. S. Baggett and L. N. Trefethen, Phys. Fl. A {\bf 9},
1043 (1997).

\bibitem{boberg} L. Boberg and U. Brosa, Z. Naturforsch. {\bf 43a}, 697 (1988).

\bibitem{farrell1} B. F. Farrell and P. J. Ioannou, PRL {\bf 72},
1188 (1994).

\bibitem{grossman} T. Gephardt and S. Grossman, Phys. Rev. E
{\bf 50}, 3705 (1994).

\bibitem{waleffe1} F. Waleffe, Phys. Fl. A. {\bf 7}, 3060 (1995).

\bibitem{holmes} P. J. Holmes, Physica {\bf 2}D, 449 (1981).

\bibitem{mccaughan}  F. E. McCaughan, SIAM J. Appl. Math. {\bf 50}, 1232
(1990).

\bibitem{guckenheimer} J. Guckenheimer and P. Holmes, {\em Nonlinear
oscillations, dynamical systems and bifurcations of vector fields}
(Springer-Verlag, New York, 1993).

\bibitem{crawford} J. D. Crawford and E. Knobloch, Physica {\bf 31}D, 1 (1988).

\bibitem{noisy} E. R. Tracy, X.-.Z Tang \& C. Kulp, ``Takens-Bogdanov random
walks'', submitted to J. Stat. Phys.

\bibitem{bagthesis} J. S. Baggett, Ph.D. Thesis, Cornell University, 1996.

\bibitem{numrec} W. H. Press, S. A. Teukolsky, W. T. Vetterling, and
B. P. Flannery, {\em Numerical Recipes}, $2^{nd}$ edition (Cambridge
University Press, New York, 1992).

\end{thebibliography}
\end{document}